\documentclass[conference]{IEEEtran}
\makeatletter
\let\@ORGmakecaption\@makecaption
\long\def\@makecaption#1#2{\@ORGmakecaption{#1}{#2}\vskip\belowcaptionskip\relax}
\makeatother 

\IEEEoverridecommandlockouts
\usepackage{cite}
\usepackage{lipsum}
\usepackage{amsmath,amssymb,amsfonts}
\usepackage{algorithmic}
\usepackage{graphicx}
\usepackage{textcomp}
\usepackage{xcolor}
\usepackage[hidelinks]{hyperref}
\usepackage[capitalize]{cleveref}
\usepackage{comment}
\usepackage{listings}
\crefname{equation}{}{}
\Crefname{equation}{Equation}{Equations}
\crefname{lstlisting}{List.}{Lists.}
\Crefname{lstlisting}{List.}{Lists.}
\def\BibTeX{{\rm B\kern-.05em{\sc i\kern-.025em b}\kern-.08em
    T\kern-.1667em\lower.7ex\hbox{E}\kern-.125emX}}
\begin{document}

\title{Large Language Model-based System to Provide Immediate Feedback to Students in Flipped Classroom Preparation Learning
}

\author{\IEEEauthorblockN{Shintaro UCHIYAMA}
\IEEEauthorblockA{\textit{Dept. of Computer Science and Engineering} \\
\textit{Toyohashi University of Technology}\\
Toyohashi, Japan \\
s183313@edu.tut.ac.jp}
\and
\IEEEauthorblockN{Kyoji UMEMURA}
\IEEEauthorblockA{\textit{Dept. of Computer Science and Engineering} \\
\textit{Toyohashi University of Technology}\\
Toyohashi, Japan}
\and
\IEEEauthorblockN{Yusuke MORITA}
\IEEEauthorblockA{\textit{Faculty of Human Sciences} \\
\textit{Waseda University}\\
Tokorozawa, Japan}
}


\maketitle
\global\csname @topnum\endcsname 0
\global\csname @botnum\endcsname 0

\begin{abstract}
This paper proposes a system that uses large language models to provide immediate feedback to students in flipped classroom preparation learning.
This study aimed to solve challenges in the flipped classroom model, such as ensuring that students are emotionally engaged and motivated to learn.
Students often have questions about the content of lecture videos in the preparation of flipped classrooms, but it is difficult for teachers to answer them immediately.
The proposed system was developed using the ChatGPT API on a video-watching support system for preparation learning that is being used in real practice.
Answers from ChatGPT often do not align with the context of the student's question.
Therefore, this paper also proposes a method to align the answer with the context.
This paper also proposes a method to collect the teacher's answers to the students' questions and use them as additional guides for the students.
This paper discusses the design and implementation of the proposed system.
\end{abstract}


\section{Introduction}
\begin{figure}[tb]
  \centering
  \includegraphics[width=\linewidth]{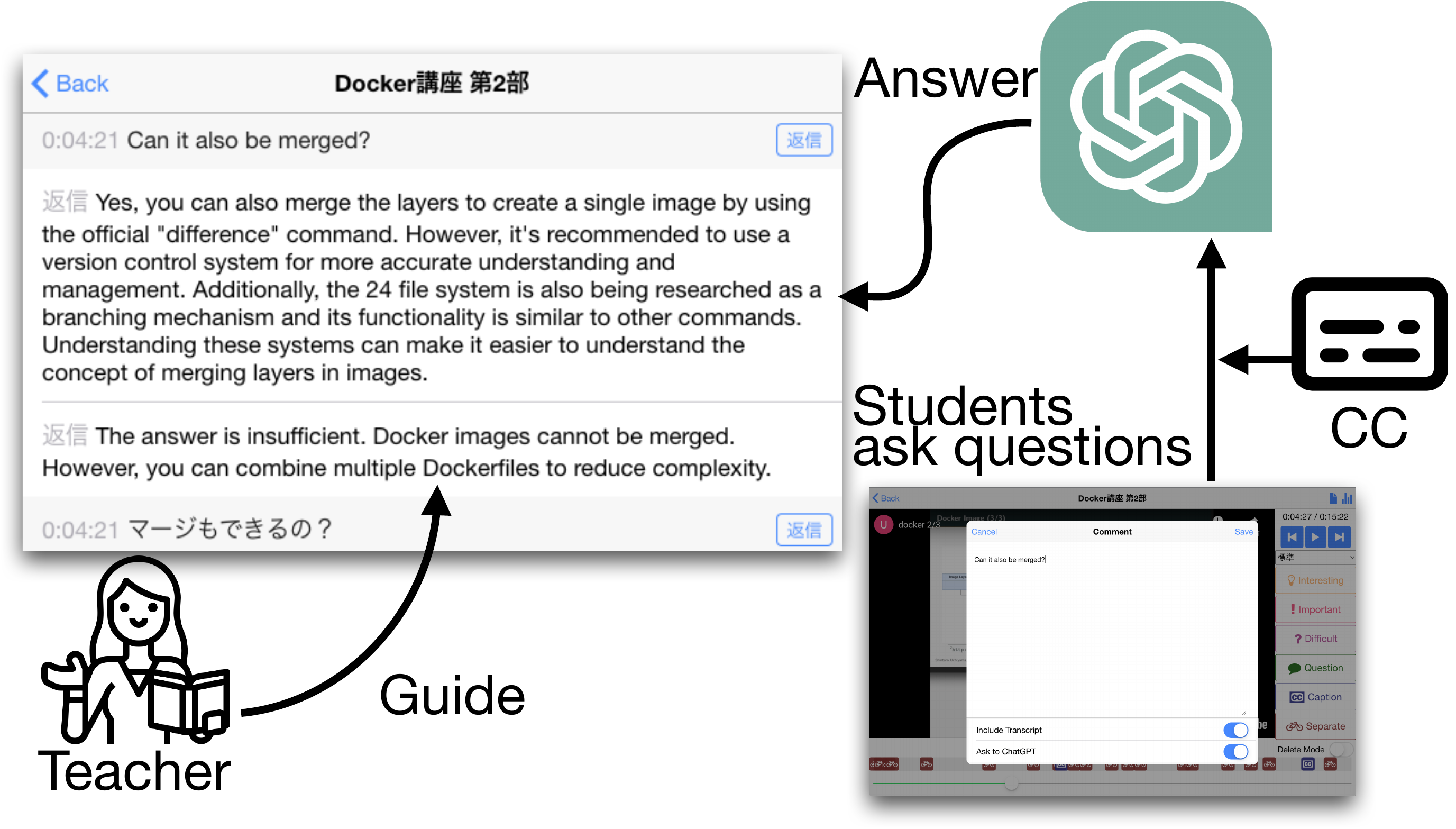}
  \caption{\textbf{Basic concept of the proposed system}: The proposed system uses a chatbot to provide immediate feedback to students in flipped classroom preparation learning. Moreover, subtitles of the lecture videos are used to specify the context of the students' questions. Teachers can also include additional guides as replies to the students' questions and their answers from a chatbot.}
  \label{fig:concept}
\end{figure}
The flipped classroom model has gained significant attention as an innovative teaching approach, aiming to increase student engagement and enhance learning outcomes\cite{Bergmann2012-lp}.
In this flipped classroom model, students learn the course content before class using videos\cite{Bishop2013-dj} and use class time to apply the knowledge they have learned.
However, despite its potential benefits, the flipped classroom model also presents several challenges, such as ensuring that students are emotionally engaged and motivated to learn\cite{Akcayir2018-sq}.
The lack of immediate feedback is an important challenge of the flipped classroom model.

One possible solution to this challenge is the use of chatbots, which can provide immediate feedback and support to students in their learning process\cite{Okonkwo2021-ak}.
Chatbots also improve students' emotional engagement and motivation to learn.
However, existing educational chatbots have limitations in their ability to provide immediate feedback to students in various educational settings.
The traditional approach to developing chatbots requires a large amount of human effort to create a knowledge base and dialogue system.

The recent advancements in large language models (LLMs) offer new opportunities for enhancing educational chatbots and their potential applications in various educational settings\cite{Kasneci2023-fr}.
LLMs can be used to generate responses to students' questions without the need for a knowledge base or dialogue system.
ChatGPT\cite{openai2022} is a famous LLM based on the GPT-3.5 model and can be easily integrated into existing systems using its API.
The use of LLMs in educational practices is still in its infancy, and there is a lack of research on how to use LLMs to improve education.
For example, the answers from LLMs often do not align with the context of curriculums and may be incorrect or inappropriate\cite{Kasneci2023-fr} if LLMs are simply used in education.
Therefore, there is a need for research on how to use LLMs to enhance educational chatbots and their potential applications in various educational settings.

Furthermore, emotional engagement plays a crucial role in students' motivation and learning outcomes in educational settings\cite{Skinner2009-qt}.
Especially in preparation learning, motivation should be treated because it can affect the learning outcomes\cite{Van_Alten2019-gl}.
Online collaboration among students affects their motivation\cite{Vezne2022-ha}.
However, not all students are good at online collaboration; some students are ``Lone Wolves'' such that they do not communicate with academic stakeholders so often\cite{Kayser2020-tn}.
``Lone Wolf'' students may only communicate with non-academic stakeholders, such as their family and friends; therefore, if there is a mechanism to acquire their relevant reactions, at least for preparation materials like through non-academic stakeholders, it would be helpful for them.
Therefore, the student's interaction with non-academic stakeholders should be known by teachers.
LLMs could act like non-academic stakeholders to improve students' emotional engagement.

This paper proposes a system to enhance students' emotional engagement in flipped classroom preparation learning using LLMs.
The basic concept of the proposed system is shown in \cref{fig:concept}.
We aim to use a chatbot to provide immediate feedback to students in flipped classroom preparation learning; however, the answer is to be treated as a tentative one.
Moreover, subtitles of lecture videos are used to specify the context of the student's questions.
By using subtitles, the chatbot can provide more accurate and relevant responses to students' questions.
This paper also discusses the challenges and future directions of the proposed system.

\section{Base System: a Video-Watching Support System for Preparation Learning}
Response Collector\cite{Okumoto2018-db}, a web application for preparation learning video watching, is used as a base system.
It collects students' annotations (referred to as responses in this paper) and behaviors for videos in sync while they watch preparation videos.

Students can annotate their responses for a video on Response Collector by just clicking a button.
The responses have four types: \textit{Interesting}, \textit{Important}, \textit{Difficult}, and \textit{Question}.
In particular, \textit{Question} is a response to asking a question about the video.
Students type their questions as text, and teachers are expected to answer them in the face-to-face class.
Teachers can also annotate their designated responses designed to guide students' learning for a video on Response Collector.
One of the designated responses is \textit{Steering Marks}\cite{Uchiyama2019-co}, a response to show the teacher's intentions of what students are expected to learn.
Another is \textit{captions} to show additional information for students on the video.

Response Collector also collects students' behaviors for videos.
It collects the time when students start watching a video, the time when they stop watching a video, and the time when they put in a response.
These behaviors can be visualized on Response Collector.

Response Collector can effectively collect and visualize students' responses and behaviors for videos in sync.
However, their feedback, especially for the question responses, is given only in the face-to-face class.
This causes a problem that students cannot get immediate feedback for their questions.
Moreover, ``Lone Wolf'' students may not ask questions because they need to get feedback in the face-to-face class, which increases their mental load.
Therefore, we improve Response Collector to provide immediate feedback from LLMs for students' questions.

\section{ChatGPT}
ChatGPT\cite{openai2022} is a conversation model based on a generative pretrained transformer (GPT), which is an LLM.
It is fine-tuned on a large dataset of conversations from resources on the Internet.
Moreover, it is trained by the human-in-the-loop method, which is a method to improve the performance of LLMs by using human feedback to make the model output more familiar to humans.
ChatGPT works very well on various tasks, such as question answering, summarization, and translation.

The input for ChatGPT is called a ``prompt''.
A prompt is a sequence of words that are given to ChatGPT to generate a response.
By changing the prompt, ChatGPT can generate various responses.
ChatGPT does not have a mechanism for long-term memory.
Therefore, the prompt must contain the whole context of the conversation.

GPT is based on the transformer, which infers the next word from the previous words.
It is trained on a huge dataset of text from the Internet.
Therefore, it just generates a word that is most likely to be the next word based on the learned tendency.
This is why the output of GPT is sometimes not appropriate for the context.
GPT simply outputs the sequence of words, so the reason why it is so powerful is still being investigated.

Some researchers have already argued that ChatGPT can be a useful tool in education\cite{Kasneci2023-fr,Bauer2023-bl,Zhai2022-bz}.
They also argued that everyone has to be careful when using ChatGPT in education because it sometimes generates inappropriate responses.
If it is simply used in education, it may cause a problem that students are given the wrong information.
In addition, due to the broad range of topics that are learned in ChatGPT, it may generate responses that are not relevant to the context of the curriculum.
Because LLMs are quite general, it is not easy to use them in education.
Also, it is not easy to include the whole context of a subject in the prompt due to the limitations of current LLMs, especially the maximum token length.
Therefore, there is a need for research on how to use LLMs effectively in education.

\section{Proposed System}
\begin{figure}[tbp]
  \centering
  \includegraphics[width=\linewidth]{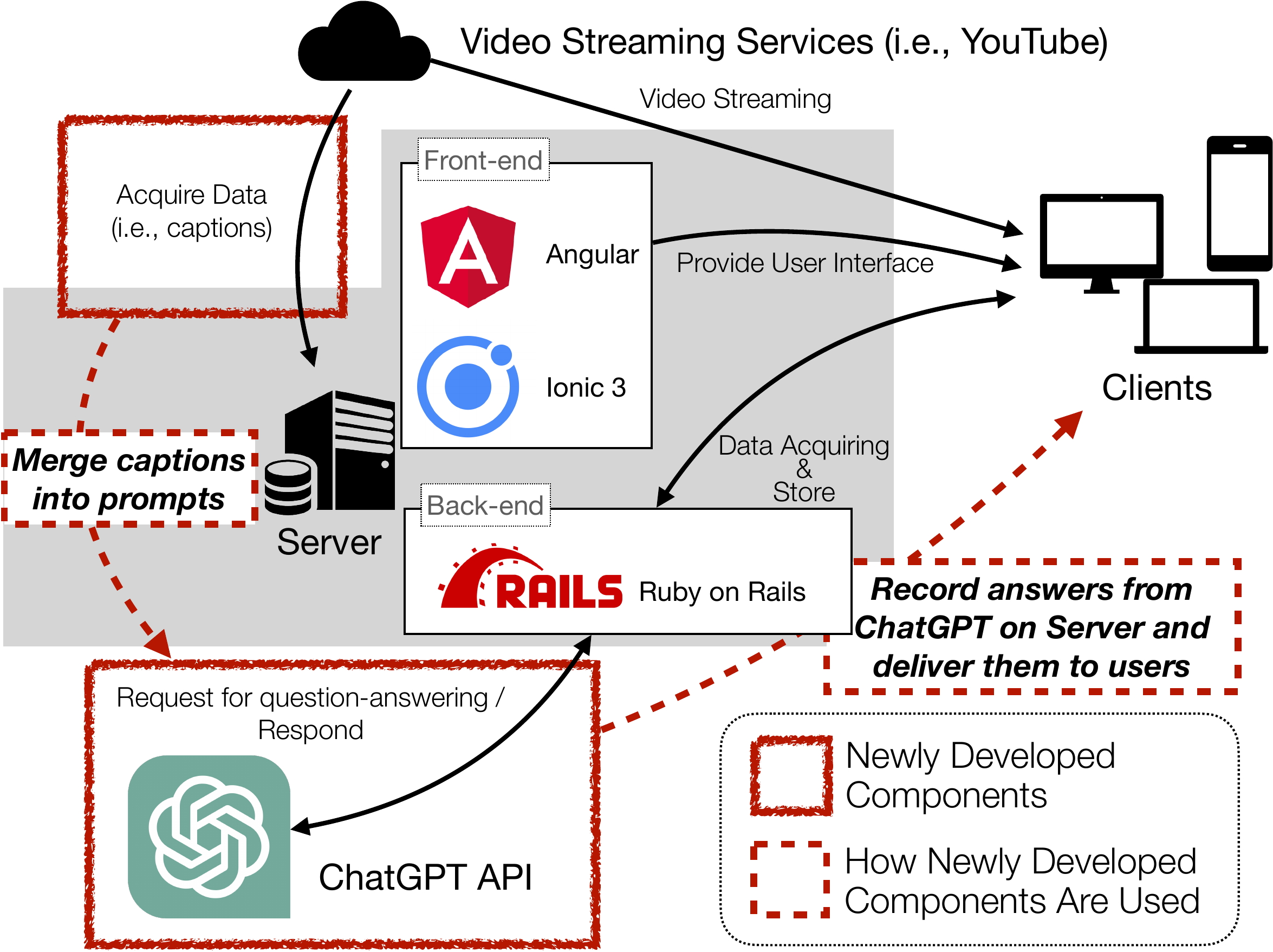}
  \caption{\textbf{Architecture of the proposed system}: The solid red boxes show newly developed components in this study. The dotted red boxes and arrows show how the components are used. The base Response Collector is made with Ruby on Rails, Angular, and Ionic; therefore, the newly developed components was also made with them. Prompts for ChatGPT include the subtitles of the lecture videos, and the answers are recorded on the server and shown to users.}
  \label{fig:arch}
\end{figure}
\begin{figure}[tbp]
  \centering
  \includegraphics[width=\linewidth]{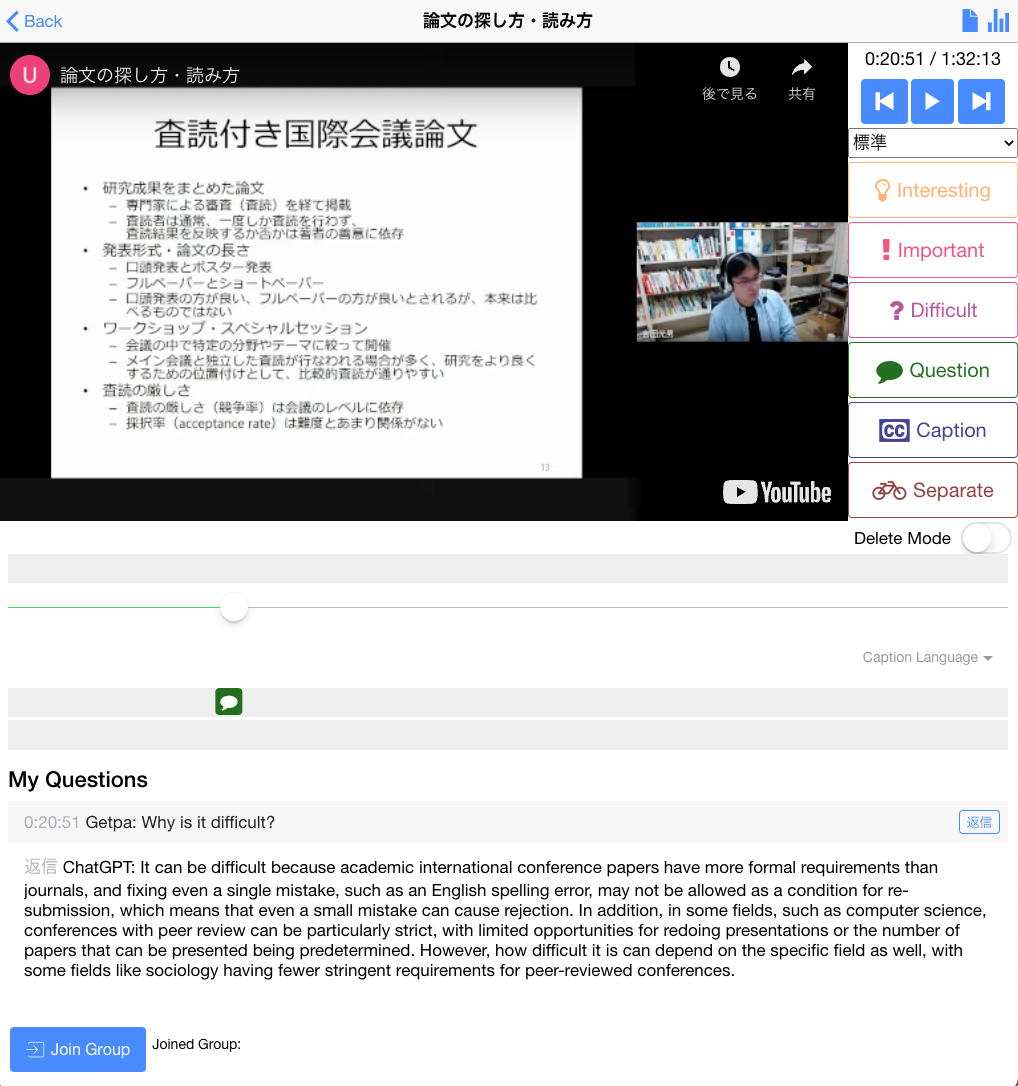}
  \caption{\textbf{Overview screenshot of the proposed system}: The bottom of the screenshot shows an example of a question and its answer from ChatGPT.}
  \label{fig:overview}
\end{figure}
\begin{figure}[tbp]
\begin{lstlisting}[
    caption={\textbf{Example of prompt} (originally in Japanese): This prompt is for a video titled \textit{What is the best usage of ChatGPT? Talking about the practical use in Education, Business, etc.}.},
    label=lst:prompt,
    belowcaptionskip=4ex
    ]
{
  "model": "gpt-3.5-turbo",
  "messages":[
    {"role": "system", "content": "You are a good teacher. As the user watches the material, he/she inputs his/her reactions to the material. If you have given the user subtitles for the material, please take them into account. The teacher responds to the user's reaction by prompting the user to think."},
    {"role": "user", "content": "### Subtitles from the video"},
    {"role": "user", "content": "I think it's fine if everyone uses AI, but principals and teachers are good at making good speeches. I think it has a low intelligence quotient and is emotionally weak, but I've been playing with it for a while now and to a certain extent I think it's better than most working people. I have also used it for user interviews in a strange way. I asked them what their problems are when they are using a blog or what they think about when they compare WordPress with something like that."},
    {"role": "user", "content": "### Question"},
    {"role": "user", "content": "I think working people are still better, how do you think?"}
  ]
}
\end{lstlisting}
\end{figure}
\begin{figure}[tbp]
  \centering
  \includegraphics[width=\linewidth]{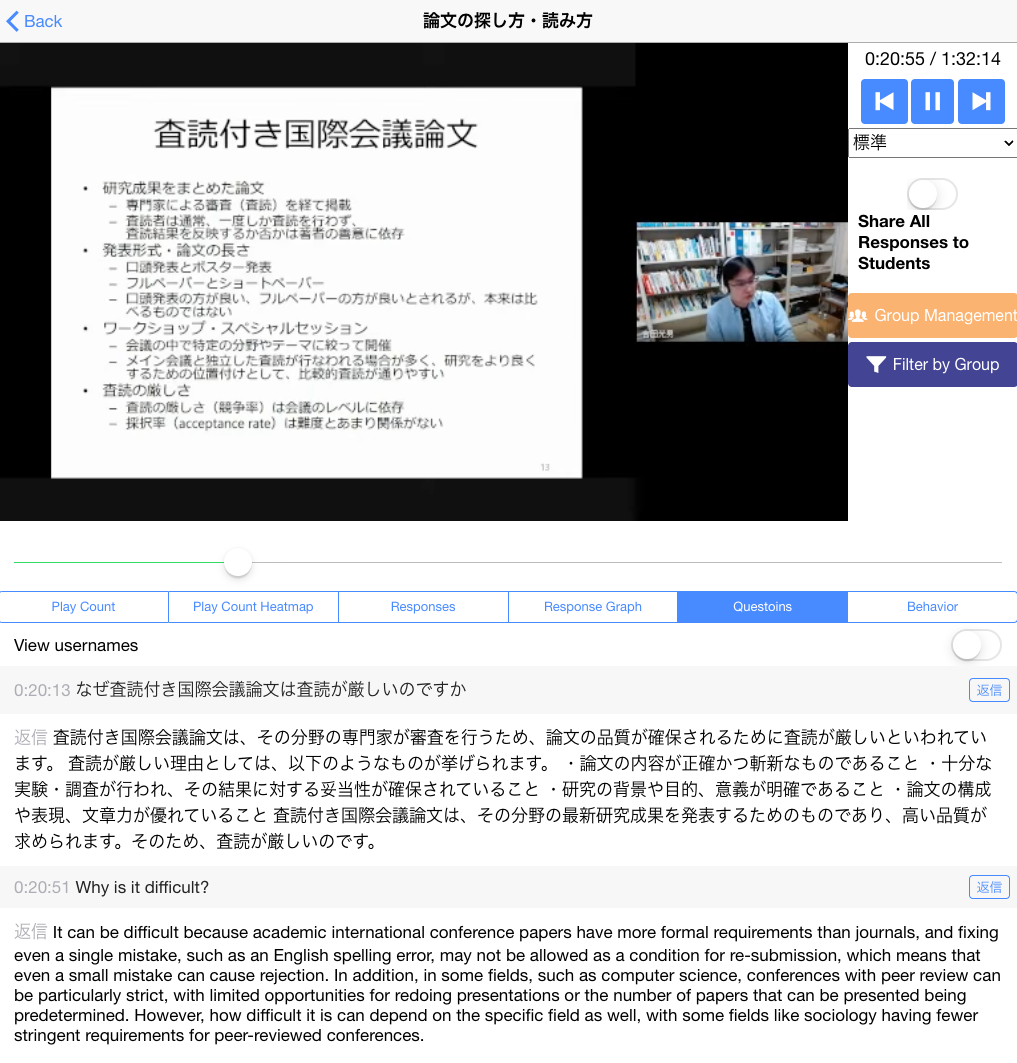}
  \caption{\textbf{Teacher's view of the proposed system}: The teacher can see all questions from students and answers from ChatGPT. The teacher can also reply to a question manually by clicking a button on the right of the question.}
  \label{fig:teacher}
\end{figure}

Our proposed system concept is shown in \cref{fig:concept}.
The concept is to use LLMs to provide immediate feedback to students in flipped classroom preparation learning, but the answer is treated as just a tentative one, so the teacher provides guides in the face-to-face class or as an additional reply.
In addition, subtitles of the lecture videos are used to specify the context of the student's questions.
According to this, the architecture of the proposed system is as  shown in \cref{fig:arch}, where the solid red boxes show the components we newly developed, and the dotted red boxes and arrows show how the components are used.
These new components were developed on the base Response Collector.

First, a function to reply to students' questions was developed.
The base Response Collector has a function to collect students' questions as responses; therefore, the function was extended to collect the answers from LLMs as replies.
This function is needed to show the answers to students' questions from the LLM.
Also, it can be used by teachers or students to simply reply to a question manually.

Then, an auto-reply function using ChatGPT\cite{openai2022} is implemented using its API as a question-answering model.
An example answer from ChatGPT is shown at the bottom of \cref{fig:overview}.
When students put a \textit{Question} response, the system sends the question to ChatGPT\footnote{Currently, we use the \texttt{gpt-3.5-turbo} model to save running costs. Of course, we could use \texttt{gpt-4} if it is available on an API key.} and gets an answer from it.
Then, the system gives the answer as a reply to the student's \textit{Question} response.
The system also records the answer on the server.
These processes are performed asynchronously so that students can get an answer in a few seconds.

Second, a subtitle retrieval function was developed to make prompt context clear.
Subtitles are retrieved from the YouTube caption server.
When students put a \textit{Question} response, the system retrieves the subtitles between the 30s before and after the video timeline of the response.
Students can switch the subtitle inclusion on and off.
If the subtitle inclusion is on, the subtitles are included in the prompt.
An example of the prompt with subtitles and the answer from ChatGPT is shown in \cref{lst:prompt}.

Lastly, the system provides a function to show all questions and their replies.
We treat the answer from ChatGPT as a tentative answer as it is not always correct.
Due to this problem, a manual reply function is required.
Teachers can see all questions and their replies on the teacher view shown in \cref{fig:teacher}.
Teachers can also reply to a question on the teacher's view.
The teacher's view is also used to show the answer to a question in the face-to-face class.

\section{Use Case}
In this section, the use case of our proposed system is described.
The system is expected to be used in flipped classrooms with preparation videos.
Let us consider one flipped classroom practice with a preparation video.

First, the teacher prepares a preparation video for the flipped classroom.
The preparation video is uploaded to YouTube, and the subtitles are added to it automatically by YouTube.
The preparation video can be made by someone other than the teacher; therefore a variety of \textit{open educational resources} (OERs) on YouTube (e.g., Khan Academy) can be used as preparation videos.
Then, the preparation video is registered to the Response Collector.

Second, students watch the preparation video on Response Collector.
Students can put their responses to the video on Response Collector.
In particular, students can put their questions about the video as \textit{Question} responses.
The questions are sent to ChatGPT, and the answers are shown as replies to the questions.
Students can also see the questions from other students in the same group and their replies on Response Collector.

Third, the teacher checks the questions and replies to the question in the teacher's view.
If the answers from ChatGPT need to be corrected or improved, the teacher can put a reply to do so.
The teacher can use the questions and replies to questions to prepare for the face-to-face class.

Lastly, the teacher reviews the questions in the face-to-face class.
Some answers from ChatGPT are expected to be wrong or insufficient.
Thus, the teacher can correct or improve the answers in the face-to-face class again.
Therefore, this creates an opportunity for face-to-face activity with students, which was not possible before.

\section{Discussion}
In this paper, we proposed a system to support flipped classrooms with LLMs.
The system is expected to be used in flipped classrooms with preparation videos.
The system has functions to reply to students' questions automatically using ChatGPT and to include the subtitles of the video in the prompt to make the prompt context clear.
These functions are expected to reduce the problems in flipped classrooms.
The proposed system was tested in a real classroom as a trial.
As a result, it was confirmed that the proposed system can be used in real classroom practice without any major issues.

Response Collector was chosen as a base system because it is already used in several lecture practices (e.g., \cite{Kobayashi2021-gm}).
This means the system has sufficient quality to be used in real lecture practices; therefore, our proposed system can be used in real lecture practices already.
In this study, as shown in \cref{fig:arch}, we developed new components without changes to existing components.
Therefore, there are no lost data or functions in the base system.
In fact, the base system is still compatible with the proposed system; thus, the base system (without LLM integration) and proposed system (with LLM integration) are available at the same time.

One of the most important challenges of preparation learning in flipped classrooms is the lack of immediate feedback.
The proposed system can provide immediate feedback to students' questions using ChatGPT.
This immediate feedback is expected to reduce students' anxiety about preparation learning.
Regarding anxiety, without immediate feedback, students may not feel like putting questions to the system.
Therefore, the function is also expected to encourage students to put more questions on the system.
The more questions students put, the more the teacher can know about students' understanding of the preparation learning.
Therefore, this function is expected to improve the quality of the face-to-face class.
Moreover, the data from the function can be used for further analysis of the students' learning.
We believe more data than the base system can be collected by the function.
Therefore, we hope that the function can contribute to the field of learning analytics by collecting more data.

``Lone Wolves'' may not make their questions if they cannot obtain immediate feedback.
Without their response, it is difficult to know their understanding of the preparation learning.
If they feel like asking the system their questions, as they ask non-academic stakeholders their questions, the proposed system could acquire responses from them.
Then, the teacher can know the student's understanding of the preparation learning, even if they are ``Lone Wolves''.
This expectation must be verified by further research.

One of the most concerning problems in using LLMs is the authenticity of their answers\cite{Kasneci2023-fr}.
The authenticity of answers from LLMs isn't guaranteed.
Therefore, answers from LLMs need to be checked for correctness by teachers or students.
The proposed system is expected to be used in flipped classrooms.
In flipped classrooms, there must be a face-to-face class after the preparation learning.
Thus, the teacher can check the answers from LLMs in the face-to-face class.
The teacher can also correct or improve the answers by replying to the questions asynchronously.
Therefore, the authenticity of the answers from LLMs is not a problem in the proposed system.

Another concerning problem in using LLMs is that the answer may not be aligned with curriculum\cite{Kasneci2023-fr}.
The curriculum reflects the use of video; therefore, the video subtitles can be treated as having the same context as the curriculum.
The proposed system has a function to include the subtitles of the video in the prompt to make the prompt context clear.
An example of the answer to the question ``I think working people are still better, how do you think?'' shown in \cref{lst:prompt} from ChatGPT \textit{without} subtitles is as follows:
\begin{quote}
I believe that being a professional and student each have their own merits and drawbacks. Being a working professional enables one to gain experience and knowledge in their career, achieve economic independence, and play a role in society. Meanwhile, being a student means that there are still limitless possibilities for the future, opportunities to gain new knowledge and skills, and time to pursue one's own potential.

Therefore, it is not accurate to say which option is better. It is important for each person to make a choice that suits them based on their environment, life plan, and goals.
\begin{flushright}
  \textit{Translated. Originally in Japanese.}
\end{flushright}
\end{quote}
This answer is not aligned with the context of the video.
The video is about ChatGPT; therefore, the answer should be about ChatGPT.
However, an example of the answer for the same question from ChatGPT \textit{with} subtitles is as follows:
\begin{quote}
I think it depends on one's perspective. Indeed, there may be areas where AI cannot compete with human abilities, such as developing human relationships and qualities.However, AI also has capabilities in processing massive amounts of data and knowledge as well as versatility that surpass human ability. Furthermore, collaboration between AI and humans can create a synergistic effect. The superiority of both depends on societal roles and needs.
\begin{flushright}
  \textit{Translated. Originally in Japanese.}
\end{flushright}
\end{quote}
This answer is aligned with the context of the video because it answers in the context of AI.
Therefore, the function is confirmed to improve the alignment of the answers from ChatGPT with the context of the video.
The function must be evaluated by further research, including considering whether it is reasonable to include the subtitles of the video in the prompt as the context and whether there are other methods.

There are some drawbacks to the proposed system.
Teachers would be required to take more time than with the base system to read the questions and answers from ChatGPT.
This would lead to an increase in preparation time for the face-to-face class.
In addition, if the answer from ChatGPT gives \textit{misconceptions}, it would be more difficult and take more time to correct the answer than simply teaching directly.
Therefore, we treat the answer from ChatGPT as just a tentative answer, and our concept shown in \cref{fig:concept} includes the teacher's guide.
Other drawbacks must be found and addressed in further research.

The proposed system must be evaluated in real classroom practice.
A questionnaire for both teachers and students to verify the acceptance of the system is being considered.
Also, analysis of the behavior and response data acquired by the proposed system is being considered.
These evaluations are planned for the near future.

\section{Conclusion}
This paper proposed a system to support preparation learning in flipped classrooms with LLMs.
The system has a function to reply to students' questions automatically using ChatGPT.
Moreover, the system has a function to include the subtitles of the video in the prompt to make the prompt context clear.
These functions are expected to reduce the problems in flipped classrooms.

This paper described the expected use case of the proposed system.
Also, it was discussed how the proposed system can solve the problems in flipped classrooms and those of using LLMs in educational practices.
We believe that the proposed system can contribute to the practice of flipped classrooms.
It can also contribute to the field of learning analytics by collecting more data.

\bibliographystyle{IEEEtran}
\bibliography{IEEEabrv,main}

\end{document}